\documentclass[twocolumn,amsmath,amssymb]{revtex4-1}
\usepackage{dcolumn}

\usepackage{hyperref}
\usepackage{graphicx}% Include figure files

\def\+{\!+\!}
\def\-{\!-\!}
\def\={\!=\!}
\begin{document}

\title{Modeling Hall viscosity in magnetic-skyrmion systems}

\author{Bom Soo Kim}
\affiliation{%
Department of Physics, Loyola University Maryland, Baltimore, MD 21210, 
USA
}%

\date{\today}

\vspace{0.5in}
\begin{abstract}
Magnetic skyrmions are topologically stable objects that are made with a bunch of spins tightly arranged in a smooth fashion. Their topological nature provides unusual and complex transport properties, such as the skyrmion Hall effect. Extensive Hall data have further revealed asymmetry between skyrmion and antiskyrmion Hall angles, which cannot be accounted by known mechanisms. Here, we explain this asymmetry by utilizing another universal transport coefficient called `Hall viscosity,' extensively studied in quantum Hall systems. Hall viscosity is modeled in steady-state skyrmions motion by generalizing the Thiele equation with a transverse velocity component and is independent of the skyrmion charge. Our analyses, based on available asymmetric Hall angle data, reveal this transverse force amounts 3\% - 5.4\% of the force due to the skyrmion Hall effect. Further clarification of Hall viscosity will be essential for designing next generation storage devices properly, not to mention for our deeper understanding of fundamental properties of nature. 
\end{abstract}

\maketitle

Skyrmions are particle-like extended spin textures that are found in magnetic materials and characterized by a topological number \cite{SkyrmionTh}\cite{SkyrmionExp1}\cite{SkyrmionExp2}. Since its discovery, magnetic skyrmion has attracted much attention due to various advantages suitable for next generation storage devices: small size, topological protection and low-energy operation cost \cite{Jonietz2010}\cite{SkyrmionTopologicalReview}. Controlling skyrmion motion is essential for the purpose. The skyrmion Hall effect has been one of the central components of recent investigations \cite{Jiang2017}\cite{Litzius2017} as it pushes skyrmion transverse to the direction of a driving force, possibly resulting in its annihilation and information loss at devices' edges. This Hall effect is sensitive to its charge, especially its sign depending on skyrmion or antiskyrmion.   

Recently, a vanishing skyrmion Hall effect has been reported at the angular momentum compensation temperature in the context of a ferrimagnet \cite{VanishingSkyrmionHall}. Among their data, we note that the skyrmion Hall angle, $\theta_{SkH}\=-35^o $, for a magnetization-up state (skyrmion) is significantly different from that of a magnetization-down state (antiskyrmion), $\theta_{ASkH}\=31^o $. Similar asymmetries can be found in the experimental data in \cite{Jiang2017}. These imbalances are unexpected based on our understanding of the skyrmion Hall effect. We offer a possible resolution here.  

Skyrmions exist in the magnetic systems with broken parity symmetry. Hydrodynamic description of such systems reveal new transport coefficients \cite{Jensen:2011xb}\cite{Bhattacharya:2011tra}, such as Hall viscosity \cite{Avron:1995}. This mysterious Hall viscosity has been vigorously investigated in quantum Hall states \cite{Read:2008rn}\cite{Hoyos:2011ez}, yet its experimental conformation is still lacking.  

Hydrodynamics solely relies on symmetries. Their transport coefficients, including Hall viscosity, are universal and expected to play crucial roles. Hall viscosity has been introduced recently to the systems with skyrmions in \cite{Kim:2015qsa}\cite{Kim:2019vxt}. Its possible experimental measurements and connections to the Hall conductivities are proposed there.  
%Hall viscosity is a universal transport coefficient that can play an important role in the system with broken parity symmetry \cite{Avron:1995} and has been introduced recently to the systems with skyrmions in \cite{Kim:2015qsa}\cite{Kim:2019vxt}. While it has been vigorously investigated theoretically in quantum Hall states \cite{Read:2008rn}\cite{Hoyos:2011ez}, its experimental conformation is still lacking.  

We study the properties of Hall viscosity using hydrodynamics and model them in Thiele equation \cite{Thiele} that describes the center of motion of a skyrmion. This generalized Thiele equation is applied to insulating and conducting magnets with skyrmions in the presence of various spin torques, such as spin transfer torque, spin orbit torque, spin Hall torque, and emergent electromagnetic fields. We show that Hall viscosity can resolve the asymmetric Hall angles between skyrmion and antiskyrmions. This transverse force due to Hall viscosity is estimated using data from two independent experiments \cite{VanishingSkyrmionHall}\cite{Jiang2017}, and, to our surprise, is a significant part of the total Hall force. Finally, we suggest clean ways to pin point the contributions due to Hall viscosity. Background materials for the entire article can be found in \cite{Kim:2019vxt}. \\

\noindent {\bf Hall viscosity \& skyrmion motion} \\ 
Universal hydrodynamics for a system with skyrmions suggests the existence of Hall viscosity $\eta_H$, anti-symmetric viscosity tensor in addition to the symmetric one \cite{Avron:1995}. For a small deformation $\xi_i$ of a fluid, a stress is produced by a strain rate, $ \dot \xi_{ij} \= \partial_i \dot \xi_j + \partial_j \dot \xi_i$, where $\dot{\;}$ is a time derivative. 
\begin{align} \label{EMTHall}
	T_{ij} & = - ( \eta^S_{ijkl} + 	\eta_{ijkl}^A )\dot \xi_{kl}  \;. 
\end{align} 
$T^{ij}$ is stress energy tensor. For a 2 dimensional fluid with rotational invariance, $ \eta^S_{ijkl} \= \eta (\delta_{ik}\delta_{jl} \+ \delta_{il}\delta_{jk} ) \+ (\zeta \! - \eta) \delta_{ij}\delta_{kl} $ and $\eta_{ijkl}^A \= -\eta_{klij}^A \= -{\eta_H}/{2} (\epsilon_{ik}\delta_{jl} \+ \epsilon_{jl}\delta_{ik} \+ \epsilon_{il}\delta_{jk} \+  \epsilon_{jk}\delta_{il})$, where $\eta_, \zeta$ and $\eta_H$ are shear, bulk and Hall viscosities and $ \delta_{ij}, \epsilon_{ij}$ are symmetric and antisymmetric unit tensors. Here, $i,j,k,l$ refer to coordinates $x,y$. 

\begin{figure}[!t]
	\begin{center}
		\includegraphics[width=0.45\textwidth]{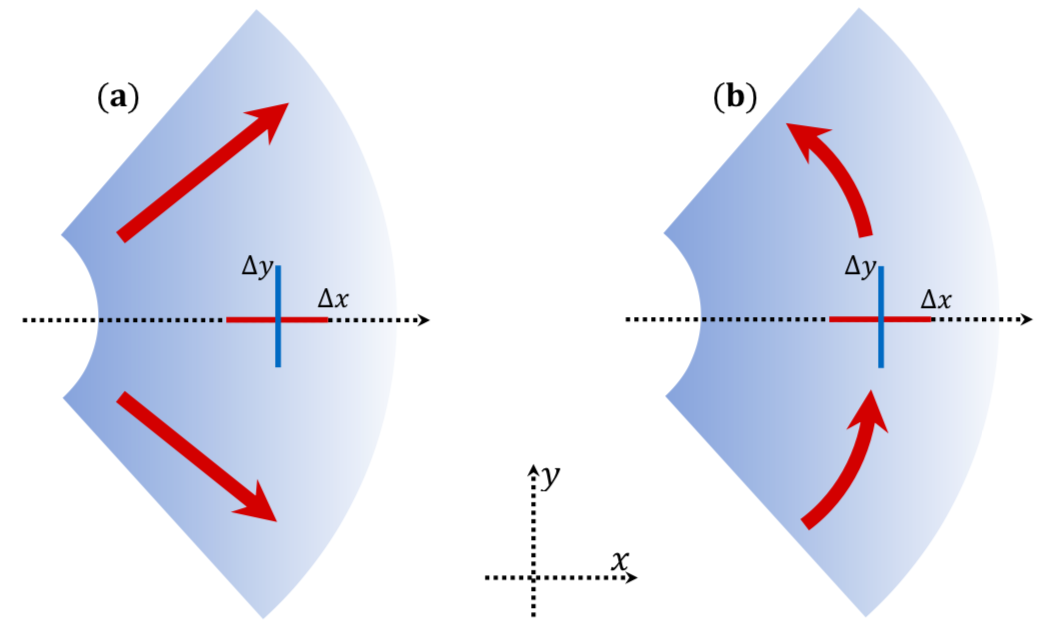} 
		\caption{\footnotesize\small Illustrations of two quadrants for rotationally invariant flows as thick red arrows. A radial flow depicted in (\textbf{a}) has $ \dot \xi_x = v_x (x)$ and $\dot \xi_y = v_y = 0$ along the $x$-coordinate, while a circular flow in (\textbf{b}) has $ \dot \xi_x = v_x =0$ and $\dot \xi_y = v_y (x)$ along the $x$-coordinate. Solid red and blue lines are small area elements $\Delta a_y \!=\! \Delta x$ facing toward $y$ direction and $\Delta a_x \!=\! \Delta y$ toward $x$ direction, respectively. \vspace{-0.2in}
		}
		\label{fig:Flow2}
	\end{center}
\end{figure}

Hall viscosity is transverse to shear viscosity. Consider a radially outward flow in Fig. \ref{fig:Flow2} (a) and compute $T^{xy} \Delta a_y$, the force due to $x$ momentum acting on small area element $\Delta a_y \= \Delta x$ facing $y$ coordinate along the horizontal direction. 
\begin{align}
	T^{xy} \Delta a_y = - 2  \eta_H (\partial_x v_x) \Delta x \;, 
\end{align}	
where we use $ \dot \xi_x\= v_x (x)$ and $\dot \xi_y\= v_y \= 0$. Note that change in $x$ momentum along $x$ direction produces force in $y$ direction via Hall viscosity. Also, the force due to $x$ momentum acting on an area $\Delta a_x\= \Delta y$ facing $x$ coordinate (in the same figure) is $T^{xx} \Delta a_x \= - 2  \eta (\partial_x v_x) \Delta y$.
These are transverse to each other. The force due to Hall viscosity acts transverse to the fluid motion and is dissipationless. This radial flow is relevant for device applications where skyrmions are pushed with pulse like forces. 

Circulating flow, curiously, shows different behaviors. From Fig. \ref{fig:Flow2} (b), $ \dot \xi_y\= v_y (x)$ and $\dot \xi_x\= v_x \= 0$. Hall viscosity is parallel to the direction of fluid motion, $T^{yy} \Delta a_y = - 2  \eta_H (\partial_x v_y) \Delta x$, while shear viscosity transverse to it, $T^{yx} \Delta a_x = - 2  \eta (\partial_x v_y) \Delta y$. These results can be confirmed in cylindrical coordinates with a finite area element. 

\begin{figure}[!t]
	\begin{center}
		\includegraphics[width=0.22\textwidth]{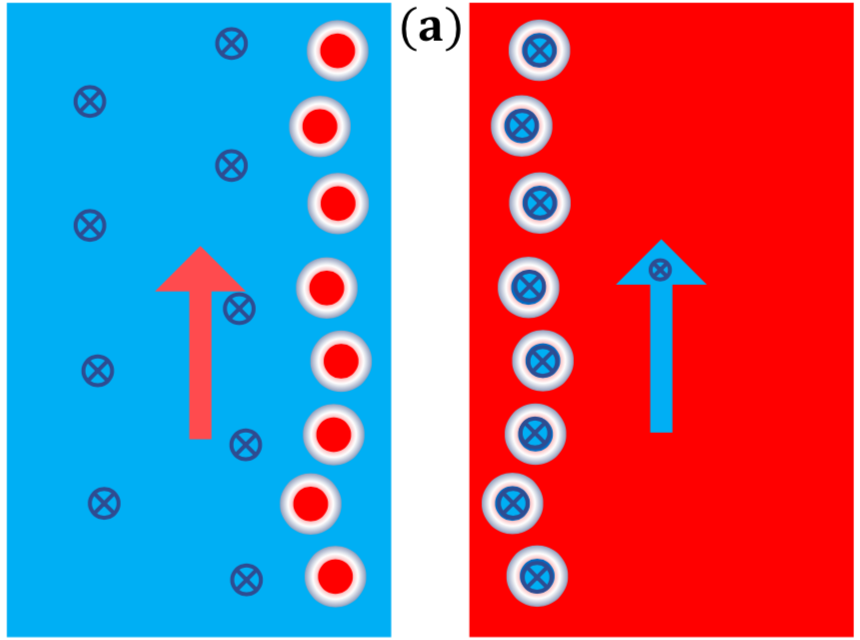} \qquad
		\includegraphics[width=0.22\textwidth]{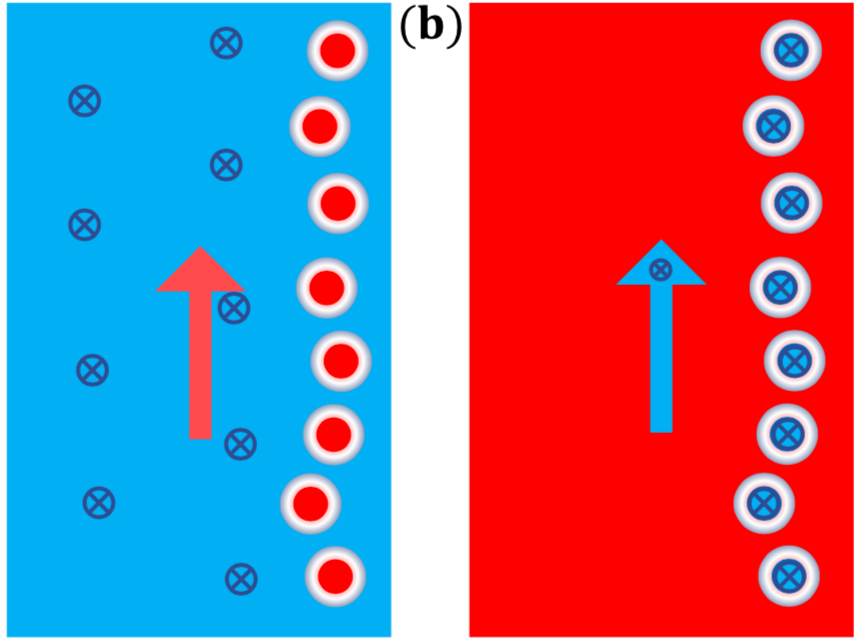} 
		\caption{\footnotesize\small (\textbf{a}) Skyrmion Hall effect: skyrmions (middle up-spin, red dots) in down-spin ferromagnet (blue background with cross circles) move toward right, while antiskyrmions (middle down-spin, blue cross circles) in up-spin ferromagnet (red background) toward left. (\textbf{b}) Hall viscosity: Both skyrmions and antiskyrmions move toward right. Thick arrow is the direction of a driving force, such as a positive electron current. Reversing the force results in turning around all figures by $180^o$.   \vspace{-0.3in} 
		}
		\label{fig:HallDrag}
	\end{center}
\end{figure}

We demonstrate that Hall viscosity and the corresponding transverse force only depend on the direction of flow and are independent of skyrmion charge. Kubo formula for Hall viscosity reads \cite{Saremi:2011ab}
\begin{align}
	\eta_H &=\lim_{\omega\to 0}\frac{\epsilon_{ik}\delta_{jl}}{4i\omega}
	\tilde G_{R}^{ij,kl} (\omega,\vec{0})\;,
\end{align} 
where $\omega$ is a frequency, repeated indices are summed over, and $\tilde G_{R}^{ij,kl}$ is a momentum space representation of the retarded Green's function for stress energy tensors $T^{ij}$, $G_{R}^{ij,kl}(t, \vec x; t', \vec x')\= -i\theta(t\- t') \langle[T^{ij}(t, \vec x),T^{kl}(t', \vec x')]\rangle$. This is an even function of the tensor $T$. Let us consider a Lagrangian, $\mathcal L \= \dot \Phi (\cos \Theta - 1) - (J/2) \partial_i \vec n \cdot \partial_i \vec n$, that accommodates skyrmions. In cylindrical coordinate $(\rho, \phi, z)$, the magnetization $\vec n$ ($\vec n^2=1$) is parametrized as 
\begin{align} \label{MagnetizationVector}
	\vec n \= (\sin \Theta (\rho) \cos \Phi (\phi), \sin \Theta (\rho)\sin \Phi (\phi), \cos \Theta (\rho) ) \;. 
\end{align}
The corresponding stress energy tensor, $T_{ij} \propto \partial_i \vec n \cdot \partial_j \vec n$, is a quadratic function of $\vec n$. We confirm that Hall viscosity is invariant under $\vec n \to \! - \vec n$, while the skyrmion charge, $Q \= \int \! d^2 x ~ q$ with $q\= \vec n \cdot ( \partial_x \vec n \times \partial_y \vec n)$, is an odd function of $\vec n$ and changes its sign. Thus, both skyrmion and antiskyrmion experience the same transverse force due to Hall viscosity as in Fig. \ref{fig:HallDrag} (b). This is contrasted to the skyrmion Hall effect depicted in Fig. \ref{fig:HallDrag} (a).  \\ \vspace{-0.1in}

\noindent {\bf Modeling Hall viscosity in Thiele equation} \\ 
Thiele described the steady state motion of a skyrmion (originally a domain wall) by parametrizing the magnetization vector $\vec n$ as $n_i \= M_i ( x_j \- X_j)/M_s$, where $M_s \= |\vec M| $ is a saturation magnetization, $x_i$ is the field position, and $X_i = v_i t$ representing the center of a skyrmion moving with a velocity $v_i$ \cite{Thiele}. 

To accommodate the transverse motion due to Hall viscosity, independent of skyrmion charge, we generalize the center of motion of skyrmion as
\begin{align} \label{SkyrmionCM}
		X_i = v_i t + R_{ij} v_j t \;,
\end{align} 
where $R_{ij} \= R \epsilon_{ij}$. The generalization does not alter the Thiele's initial setup: (i) conserved magnetization, $M_i \partial_t M_i \=  0$ and thus $|\vec M| \= const. $, and (ii) spatially constant saturation magnetization, $ M_i \partial_j M_i \= 0$. The former is a consequence of Landau-Lifshitz-Gilbert (LLG) equation, while the latter is followed by the parametrization $ M_i ( x_j \- X_j) $ and \eqref{SkyrmionCM} for $\partial_t M_i \= (v_j + R_{jk} v_k) \partial_j M_i $. 

From the fact that $M_i$ and $\partial_t M_i$ are orthogonal, Thiele showed the equation 
\begin{align} \label{LLGEQ}
	 - \frac{\epsilon_{jkl} M_k \partial_t M_l}{\gamma_0 M_s^2}  \- \alpha \frac{ \partial_t M_j }{\gamma_0 M_s} \+ \tilde \beta M_j \+ H_j = 0 
\end{align}
is equivalent to LLG equation. One can check it by multiplying $ -\epsilon_{jik} M_k$ to \eqref{LLGEQ}, summing over $j$, and renaming the indices. Multiplying $M_j$ with \eqref{LLGEQ} fixes the value $\tilde \beta \= - M_j H_j/M_s^2$, which does not contribute below. $ \gamma_0$ is the gyromagnetic ratio and $\alpha$ is a damping parameter. 

By multiplying $-\partial M_j/ \partial x_i $ on \eqref{LLGEQ} followed by integrating over unit skyrmion volume, one arrives at the generalized Thiele equation. 
\begin{align} \label{ThieleEQ}
	\mathcal G_{ij} (v_j + R_{jk} v_k) \+ \alpha \mathcal D_{ij} (v_j + R_{jk} v_k) \+ F_i = 0 \;,
\end{align} 
here $i,j,k\=x, y$ in \eqref{ThieleEQ},  
\begin{align}
	&\mathcal G_{ij}\=\epsilon_{ij} Q (M_s/\gamma_0) \;, \\
	&\mathcal D_{ij} \= -(M_s/\gamma_0)  \large\int d^2 x (\partial_i n_k) (\partial_j n_k) \;,
\end{align}	
where $\mathcal G_{ij} v_j$ is the Magnus force with the skyrmion charge $Q$, $\mathcal D_{ij} v_j$ is total dissipative drag force, and $ F_i \= -M_s \int d^2 x (\partial_i n_j) H_j$ is total external force that can include forces due to various spin torques. Internal forces due to anisotropy and exchange energies, internal demagnetizing fields, magnetostriction do not contribute \cite{Thiele}. Neither the third term in \eqref{LLGEQ}. 

We consider a skyrmion configuration $\vec n$ parametrized by $\Phi (\phi) \= m\phi + \delta_0$, with the integer $m$ characterizing the topological skyrmion charge. Then, $Q\=4\pi m$,  $\mathcal D_{xx} \= \mathcal D_{yy}  \=-(M_s/2 \gamma_0) \!  \int \! \rho d\rho (\Theta'(\rho)^2 + m^2 \sin^2 \Theta/\rho^2) $, and $\mathcal D_{xy} \= \mathcal D_{yx} \propto \int_0^{2\pi} d\phi \sin (2\phi) \= 0 $. We check the drag term $\mathcal D_{ij} v_j \!\equiv\! (M_s/\gamma_0) \mathcal D v_i$ is parallel to a force direction $F_i$. A new contribution, 
\begin{align} \label{HallViscosityContributionI}
	\mathcal D_{ij} R_{jk} v_k = (M_s/\gamma_0) \mathcal D R \epsilon_{ik} v_k \;, 
\end{align}
is transverse to $F_i$. It is quadratic in $\vec n$ and independent of the skyrmion charge. The physical origin of this {\it universal transverse drag force} can be only identified as Hall viscosity discussed above. There exist other possible transverse forces surveyed and studied below, but they are from various spin torques depending on applied currents and can be distinguished easily. 

Generalization \eqref{SkyrmionCM} adds another contribution in \eqref{ThieleEQ}, 
\begin{align} \label{HallViscosityContributionII}
	\mathcal G_{ij} R_{jk} v_k \= - Q R (M_s/\gamma_0)  v_i \;,
\end{align}
that depends on the skyrmion charge. Note it is parallel to the forcing direction. The combination $QR$ from the skyrmion Hall effect and Hall viscosity accelerates skyrmion and decelerates antiskyrmions, or vice versa, depending on the sign of $R$. This is in contrast to the transverse force \eqref{HallViscosityContributionI}. Thus, generalized Thiele equation \eqref{ThieleEQ}, that includes \eqref{HallViscosityContributionI} and \eqref{HallViscosityContributionII}, describes drastically different Hall transport phenomena.
 \\ \vspace{-0.1in}

\noindent {\bf Generalized the skyrmion Hall angle} \\ 
We consider the generalized Thiele equation \eqref{ThieleEQ} and set $F' \= (\gamma_0/M_s)  F_x $ and $F_y \= 0$ for simplicity. 
\begin{align} \label{VelocityEQ}
	(\alpha \mathcal D - QR) v_x &+ (Q+ \alpha \mathcal D R) v_y + F' = 0 \;,  \\
	-(Q + \alpha \mathcal D R) v_x &+ (\alpha \mathcal D - Q R) v_y = 0 \;. 
\end{align}	  
One gets $v_x \= (QR  - \alpha\mathcal D )  F' /(Q^2+\alpha^2 \mathcal D^2) (1+R^2)$ and $v_y \= - (Q + \alpha\mathcal D R) F' /(Q^2+\alpha^2 \mathcal D^2) (1+R^2)$. Velocity along the forcing direction is determined by $\alpha \mathcal D - QR $, which depends on the combination $QR$ mentioned in \eqref{HallViscosityContributionII}, while the transverse velocity determined by $ Q + \alpha\mathcal D R $.   

The generalized skyrmion Hall angle is 
\begin{align} \label{SkHangle}
	\tan \Theta_{SkH} = \frac{v_y}{v_x} =\frac{Q + \alpha\mathcal D R}{\alpha\mathcal D - QR } \;. 
\end{align}	
Note generalized Hall angle $\Theta_{SkH}$ depends on $R$ as well as $Q$. $\Theta_{SkH}$ reduces to the usual the skyrmion Hall angle $\tan \theta_{SkH} \= Q/\alpha\mathcal D $ when $R\=0$ in the absence of transverse drag term, and antiskyrmion has the same Hall angle as skyrmion, $\theta_{ASkH} \= -\theta_{SkH}$. Eq. \eqref{SkHangle} is useful for insulating and conducting magnets with (Anti-)skyrmions with combined applied forces in one direction, chosen to be $x$ coordinate here.    

Hall angle measurements have been performed on N\'eel-type magnetization up and down domain walls (identified as half skyrmions and antiskyrmions) in ferrimagnetic GdFeCo/Pt films by utilizing the spin orbit torque (SOT) \cite{VanishingSkyrmionHall}. These states revealed significantly different Hall angles, which prompt us to use $\Theta_{SkH}$ instead of $\theta_{SkH}$. Details of forcing mechanism are not important as $F'$ cancels out in \eqref{SkHangle}. Moreover, our analysis below shows that SOT does not have significant effects on the transverse motion of the N\'eel-type skyrmions. 

We estimate $R$ in \eqref{SkHangle} from experimental data in \cite{VanishingSkyrmionHall}. Hall angle for magnetization-up states (half skyrmion $Q\=1/2 \!\times\! 4\pi \= 2\pi$) is measured as $\Theta_{SkH}\=-35^o $, while that for magnetization-down states (half antiskyrmion $Q\=-2\pi$) is $\Theta_{ASkH}\=31^o $. Using \eqref{SkHangle} twice, $R\= -0.035$ and $\alpha \mathcal D \= -9.68$. We also compute the ratio of transverse force due to Hall viscosity $\alpha \mathcal D_{ij}  R_{jk} v_k$ to that due to the skyrmion Hall effect $\mathcal G_{ij} v_j$. 
\begin{align} \label{HVEstimate}
	\frac{\text{Hall viscosity}}{\text{Skyrmion Hall effect}} 
	\= \frac{\alpha \mathcal D R}{Q} = 5.4\% \;.
\end{align}	
It is a significant contribution. For this, we assume that the universal transverse force (independent of skyrmion charge) is entirely from Hall viscosity. We also assume that possible pinning effects due to material imperfection, excitations of internal degrees of freedom, and shape distortions for skyrmions are the same for antiskyrmions, along with system specific forces such as tension presented in \cite{VanishingSkyrmionHall}. We also recount the result \eqref{HVEstimate} by surveying various other spin torques, especially SOT below.  \\  \vspace{-0.1in}

\noindent {\bf Thiele equation with various spin torques} \\ 
Thiele equation \eqref{ThieleEQ} has been generalized with forces from different spin torques, such as spin orbit torque, spin Hall torque, spin transfer torque, and emergent electromagnetic fields. One or a combination of them play a dominant role on skyrmion dynamics depending on physical situations. Force, $F_i = \int d^2 x f_i $, is integral of force density $f_i$ over the skyrmion unit volume.

{\it Spin orbit torque} (SOT): Local magnetization in uniform ferromagnets experiences a torque through indirect $s$-$d$ exchange interactions with conduction electron spins, which feel the so-called Rashba magnetic field in their rest frames due to local electric field \cite{SOT1}\cite{Litzius2017}. 
There are two independent damping-like, $f^{SOT}_{i, DL}$, and field-like, $f^{SOT}_{i, FL}$, SOT force densities with corresponding parameters $a$ and $a \tilde \eta$. 
\begin{align} \label{ThieleEQSOT}
	&f^{SOT}_{i, DL} \= a \epsilon_{zlj} \epsilon_{lpq} (\partial_i n_p) n_q J^e_j \;, \\
	&f^{SOT}_{i, FL} \= -a \tilde \eta \left[ \epsilon_{zpq} n_j \-  \epsilon_{jpq} n_z  \right] n_p (\partial_i n_q) J^e_j  \;. 
\end{align}
Note they depend linearly on the electric current $\vec J^e$. We choose $J^e_x\= J^e, J^e_y\= 0 $ and use the same parametrization $\Phi(\phi)\= m\phi +\delta_0$ as in  \eqref{HallViscosityContributionI} for the magnetization vector $\vec n$. 
 
Damping-like $F^{SOT}_{i, DL}$ have their force components as 
\begin{align} \label{ThieleEQSOTDL}
	&F^{SOT}_{x, DL} \=  \delta_{m,\pm 1} C_{a}  \pi \cos (\delta_0) J^e \;, \\
	&F^{SOT}_{y, DL} \=  \delta_{m,\pm 1} C_{a}  m \pi \sin (\delta_0) J^e \;, \label{ThieleEQSOTDL2}
\end{align}	
where $C_{a} \= -a \int \! d\rho (\rho \Theta' (\rho) \+ \sin 2\Theta (\rho) /2 )$. N\'eel-type skyrmions ($\delta_0 \= 0, \pi $) do not receive transverse forces, while there are longitudinal forces for $m=\pm 1$. Thus damping-like SOT force can be added to $F'$ and does not play a role in the estimate \eqref{HVEstimate}. We mention that Bloch-type skyrmion ($\delta_0 \= \pm \pi/2 $) has a non-vanishing transverse force component.   

Field-like SOT force, integrated over the skyrmion volume, gives $F^{SOT}_{i, FL} \propto \rho \sin \Theta (\rho) |_{\rho=0}^{\rho=\infty}$ and vanishes for ideal skyrmions as $\sin \Theta (0) \= \sin \Theta (\infty) \=0 $. While field-like SOT can play important roles in Hall transport when skyrmion shape is deformed \cite{Litzius2017}, numerical analysis in \cite{VanishingSkyrmionHall} shows that it does not alter the main results significantly. These analysis confirms that the estimate \eqref{HVEstimate} is reasonable after taking into account of the full SOT forces. 

{\it Spin Hall torque} (SHT): When a ferromagnetic layer is placed on top of a heavy metal layer, polarized electric currents along the heavy metal layer $\vec J^{HM} $ can be used to pump polarized spins into the ferromagnetic layer through the spin Hall effect \cite{Hirsch1999}\cite{Jiang2017}. The Thiele equation has the corresponding force term  
\begin{align} \label{ThieleEQSHT}
	f^{SHT}_i \=  -b \epsilon_{zlj} \epsilon_{lpq} (\partial_i n_p)n_q  J^{HM}_j \;, 
\end{align}
where $b$ parametrizes the strength of the SHT that depends on spin Hall angle. We notice \eqref{ThieleEQSHT} has the same structure as the damping-like SOT force given in \eqref{ThieleEQSOT}. N\'eel-type skyrmions do not receive transverse forces.  

Hall effect for N\'eel-type skyrmion has been measured using SHT with pulsed high current in Ta/CoFeB/TaO$_x$ material that has strong pinning potential due to randomly distributed defects \cite{Jiang2017}. The following table collects four sets of `saturated' skyrmion Hall angle data for opposite charges $\pm Q\=\pm 4\pi$ and currents $\pm J_e $ from figure 3{\bf c} of \cite{Jiang2017}. $B$ is applied magnetic field in Oe unit. Their data were collected for different magnetic field, on which the value of the skyrmion Hall angle depends. Thus we take a linear fit to normalize the Hall angle at $B=\pm 5.0$ Oe, which is in the last column.    

\begin{table}[!h]
\begin{center}
	\begin{tabular}{|c || c | c | c || c | } 
		\hline
		   &  $(J_e,Q)$ & $(B,\Theta_{(A)SkH})$ & $(B,\Theta_{(A)SkH})$ & $(B,\Theta_{(A)SkH})$   \\ 
		\hline\hline   
		I &  $(+,-)$  & (4.8, 28$^o$) &  (5.4, 32$^o$) & (5.0, 29.3$^o$)  \\ 
		II &  $(-,+)$ & (-4.6, 28$^o$) & (-5.2, 34$^o$) & (-5.0, 32$^o$) \\ 
		III &  $(-,-)$ & (4.8, -29$^o$) & (5.4, -33$^o$) & (5.0, -30.3$^o$) \\ 
		IV &  $(+,+)$ & (-4.6, -29$^o$) & (-5.2, -33$^o$) & (-5.0, -31.6$^o$) \\  
		\hline
	\end{tabular}
\end{center}
\end{table} \vspace{-0.15in}

We note $|\Theta_{SkH}| \= 31.6^o $ (IV) and $\Theta_{ASkH} \= 29.3^o$ (I) are significantly different ($\Delta |\Theta|\= 2.3^o $) for $\+J_e$ and $B\=\pm 5.0 $ Oe. Similarly, $\Delta |\Theta|\= 1.7^o $ for $\-J_e$. The directions illustrated in Fig. \ref{fig:HallDrag} are consistent with these experimental results for $\+J_e$. Left figure in Fig. \ref{fig:HallDrag} (a) (illustrating IV) has a bigger Hall angle than the right one (illustrating I). Adding the left ones of Fig. \ref{fig:HallDrag} (a) and (b) with appropriate magnitudes produces IV, while adding the right ones does I. The other sets for $-J_e$ are also consistent. 

We use \eqref{SkHangle} (angles measured with respect to the force direction) for I and IV to get $R\=-0.020 $, $\alpha \mathcal D \= -1.70 $ and the Hall viscosity to the skyrmion Hall effect ratio $\alpha \mathcal D R/Q \= 3.5 \%$. The other data sets, II and III, give $R\=-0.015 $, $\alpha \mathcal D \= -1.65 $, and $\alpha \mathcal D R/Q \= 2.5 \%$. By taking their average, 
\begin{align} \label{HVEstimate2}
	\frac{\text{Hall viscosity}}{\text{Skyrmion Hall effect}} \= \frac{\alpha \mathcal D R}{Q} = 3\% \;.
\end{align}	

We also note the skyrmion Hall angles are different when current is reversed. The corresponding Hall angle changes, $\Delta |\Theta|\= 1^o $ for $-Q$ and $\Delta |\Theta|\= 0.33^o $ for $Q$, are insignificant compared to those due to reversing the skyrmion charge. This is consistent with the experimental results in \cite{Litzius2017}, where the skyrmion Hall angles driven by opposite currents are shown to be the same. Various `not-saturated' skyrmion Hall angle data presented in \cite{Jiang2017}, figures 2{\bf r} and 2{\bf x} for example, provide higher Hall viscosity to Hall skyrmion effect ratio than \eqref{HVEstimate2}. 

{\it Spin transfer torque (STT)}: Spins of conduction electrons moving with velocity $\vec v_s$ interact with local magnetization. This spin transfer torque is sensitive to the spatial variation of the magnetization \cite{ZhangLi}\cite{Thaiville}. 
The spatial variation of magnetization, $(v^s_i \partial_i) \vec M $, fits nicely together with $ \partial_t \vec M \propto (v_i \partial_i) \vec M $ in LLG equation. This gives two additional contributions to Thiele equation \eqref{ThieleEQ} as
\begin{align} \label{ThieleEQSTT}
	\mathcal G_{ij} (v_j\! \- v_j^s\! \+ R_{jk} v_k \!) \+ \alpha \mathcal D_{ij} (v_j \- \frac{\beta}{\alpha} v_j^s \! \+ R_{jk} v_k \!) \=- F_i \;,
\end{align} 
where $\beta$ is the non-adiabatic parameter. 

We set $F' \= (\gamma_0/M_s)  F_x $, $v_x^s\= v^s$, $F_y \= v_y^s\=0$ to get  
\begin{align} \label{VelocityEQSTT}
	&(\alpha \mathcal D \- QR) v_x \+ (Q\+ \alpha \mathcal D R) v_y \= 	(\beta \mathcal D \- QR) v^s \-  F' \;,   \\
	&(Q \+ \alpha \mathcal D R) v_x \- (\alpha \mathcal D \- Q R) v_y \= (Q \+ \beta \mathcal D R) v^s \;. 
\end{align}	  
This reduces to \eqref{VelocityEQ} for $v^s\= 0 $. Generalized skyrmion Hall angle is modified. 
\begin{align} \label{SkHangleSTT}
	\tan \Theta_{SkH} \=\frac{(Q \+ \alpha\mathcal D R)F' \+ (\alpha \- \beta) Q\mathcal D (1\+ \!R^2) v^s}{(\alpha\mathcal D \- QR)F' \! \-  (Q^2\+ \alpha\beta \mathcal D^2 ) (\! 1\+ \!R^2) v^s } \;.  
\end{align}	
Note that $\Theta_{SkH}$ depends on $F'$ and $v^s$ and is useful to model the skyrmion Hall angle when there are offsets between them. If $v_x^s \! \propto\! F' $, $\Theta_{SkH}$ is independent of $F'$. More general cases are easy to work out. 

Lastly, we include the force contribution of emergent electromagnetic fields \cite{FerroCoupling2}\cite{DMinteractionLGModel} to Thiele equation \eqref{ThieleEQ}. 
\begin{align} \label{ThieleEQEM}
	f^{EM}_i \=  Q_2 ( \beta v_i^s - \alpha v_i - \alpha R_{ij} v_j )   \;,
\end{align}
where $Q_2 \= \int d^2 x~\! q^2/\gamma_0 $ with the topological charge density $q$ defied around \eqref{MagnetizationVector}. $Q_2$ is independent of the sign of the topological skyrmion charge. For $v_x^s\=v^s$ and $v_y^s=0$, Thiele equation \eqref{ThieleEQ} becomes \eqref{VelocityEQ} with $ F'\= -\beta Q_2 v^s$ and $ \mathcal D \to \mathcal D -Q_2$. The Hall angle $\tan \Theta_{SkH}$ in \eqref{SkHangle} is modified accordingly. As the size of damping is expected to increase with this new contribution, the Hall angle will decrease, while the ratio between the transverse force due to the Hall viscosity and the skyrmion Hall effect, given in \eqref{HVEstimate}, increases. Thus, our estimates \eqref{HVEstimate}\eqref{HVEstimate2} are understated.  \\   \vspace{-0.1in}

\noindent {\bf Outlook} \\
The generalized Thiele equation can be used to investigate the existence of Hall viscosity more effectively in different settings. First, skyrmions in anti-ferromagnetic materials experience the transverse force due to Hall viscosity even in the absence of conventional skyrmion Hall effect. Eq. \eqref{VelocityEQ}, with $Q=0$, gives Hall viscosity angle, 
\begin{align}
	 \tan \Theta_R \= R \;,
\end{align}
moving toward right with respect to force direction. Note this angle should be the same for skyrmions and antiskyrmions for it is independent of the skyrmion charge. Experimental verifications would be straightforward.  \\

Second, in the presence of Hall viscosity, the precise location of vanishing skyrmion Hall effect in various parameter spaces, if any, would be deviated from the expected one. Resolving the mismatch will provide more effective and precise experimental verifications of Hall viscosity.   

In this article, we model the effects of Hall viscosity by generalizing Thiele equation with the transverse velocity component $R_{ij} v_j $. We compute the skyrmion Hall angle and examine the physical consequences due to the asymmetry of skyrmion and antiskyrmion Hall angles. According to our estimates, universal Hall viscosity produce transverse force about 3\% - 5.4\% of the skyrmion Hall effect due to topological skyrmion charge. Experimental verification of Hall viscosity will be essential for designing next generation storage devices, not to mention for understanding fundamental properties of nature.

\end{document}